\documentstyle[prl,aps,multicol]{revtex}
\input{epsf}
\begin{document}
\draft
\title{Asymptotic Statistics of Poincar\'e Recurrences\\
in Hamiltonian Systems with Divided Phase Space }                      

\author{B. V. Chirikov and D. L. Shepelyansky}
\address {Laboratoire de Physique Quantique, UMR 5626 du CNRS,
Universit\'e Paul Sabatier, F-31062 Toulouse Cedex 4, France}
\address{and Budker Institute of Nuclear Physics, 630090 Novosibirsk, Russia}

\date{\today}
\maketitle
\begin{abstract}
By different methods we show that for dynamical chaos in the standard map
with critical golden curve the Poincar\'e recurrences $P(\tau)$
and correlations $C(\tau)$ asymptotically decay in time as
$P \propto C/\tau \propto 1/\tau^3$. It is also explained why this
asymptotic behavior starts only at very large times.
We argue that the same exponent $p=3$
should be also valid for a  general chaos border.
\end{abstract}

\begin{multicols}{2}
\narrowtext

During last two decades the local structure of phase space
in chaotic hamiltonian systems and area-preserving maps had been
studied in great detail \cite{green,mackay,kadanof,meiss,stark}.
These researches allowed to understand the universal scaling
properties in a vicinity of critical invariant curves
where coexistence of chaos and integrability goes on smaller and smaller
scales in the phase space.
The most studied case is the critical golden curve
with the rotation number $r_g=[111...]=(\sqrt{5}-1)/2$
for which the scaling exponents were found with high 
precision and it was shown
that the phase space structure is self-similar and
universal \cite{mackay}. The most studied map with mixed
integrable and chaotic components 
is the standard map
\cite{chirikov} where the golden curve is critical
at the chaos parameter $K=K_g=0.97163540631$ \cite{mackay}. 
It is believed that
for $K>K_g$ all invariant Kolmogorov-Arnold-Moser (KAM) 
curves are destroyed \cite{mackay}.

While the local structure of divided phase space is now well understood,
the statistical properties of dynamics still
remain unclear in spite of simplicity of these systems.
Among the most important statistical characteristics are
the correlation function decay in time $C(\tau)$ and the statistics of
Poincar\'e recurrences $P(\tau)$. The later is especially convenient
for numerical simulations due to its natural property $P(\tau) > 0$
and statistical stability.
The first studies of $P(\tau)$ in a separatrix map showed 
that at a large time the recurrences decay as a power law
$P(\tau) \propto 1/\tau^p$ with the exponent $p \approx 1.5$
\cite{kiev}. Investigations of other different maps also indicated 
approximately the same value of $p$ \cite{karney,chsh} even if it
was remarked that $p$ can vary from map to map,
and that the decay of $P(\tau)$ can even oscillate with
$\ln \tau$ \cite{kiev,karney,chsh,dubna}. Such a slow decay of
Poincar\'e recurrences was attributed to the sticking
of trajectory near a critical KAM curve which restricts
the region of chaos \cite{kiev,karney,chsh,dubna,chispan,chilos}.
Indeed, when approaching the critical curve with the border
rotation number $r_b \;$, the local diffusion rate $D_n$ goes to zero as
$D_n \sim |r_b-r_n|^{\alpha/2} \sim 1/q_n^{\alpha}$ 
with $\alpha=5$ \cite{chispan} where 
$r_n = p_n/q_n$ are the convergents for $r_b$
determined by continued fraction expansion. The theoretical
value  $\alpha=5$ was derived from a resonant theory 
of critical invariant curve \cite{chispan,chilos}
and was confirmed by numerical measurements of local diffusion
rate in the vicinity of critical golden curve in the
standard map \cite{ruffo}. Such a decrease of the diffusion rate
near the chaos border would give the exponent $p=3$ if to assume
that everything is determined by the local properties of 
principal resonances $p_n/q_n$ given by
convergents of $r_b$
\cite{chispan,chilos,handson,dubna}. However, the value $p=3$ 
is strongly different from the numerically found 
$p \approx 1.5$. Moreover, the special simulations for the standard 
and separatrix maps with the border rotation number $r_b = r_g$
have given different behavior of $P(\tau)$ and different $p$ 
\cite{kiev,dubna} in spite of the fact that the local structure 
of the golden critical curve is universal. Different attempts were
done to reserve this difficulty. In \cite{ott} the authors argued
that a contribution from non principal resonances can reduce the  
exponent down to $p=2$. Other arguments based on disconnection of principal
resonance scales were proposed in \cite{dubna}, while Murray
discussed a possibility that larger times are required to see
$p=3$ decay \cite{murray}. During these years different Hamiltonian
systems were studied where the values of $p \approx 1 - 2.5$
have been found \cite{chsh,geisel,phystod,zaslav,artuso}. 

The analysis of Poincar\'e recurrences is interesting not only by itself but
also because they are directly related to the correlation 
function of dynamical variables 
\cite{karney,chsh,dubna,chispan,chilos,ruffo,handson,ott}:
\begin{eqnarray} 
\label{corel}
C(\tau) \sim \mu(\tau) \sim \tau P(\tau)/<\tau> \sim 1/\tau^{p-1}
\end{eqnarray}
Here $\mu(\tau)$ is the probability to remain in a given region
for time $t > \tau$, proportional to its measure,
and $<\tau>$  is the average recurrence time. This relation can be understood 
as follows. According to the definition $P(\tau)=N_{\tau}/N$
where $N$ is the total number of recurrences and $N_{\tau}$
is the number of recurrences with time $t>\tau$. 
Therefore, for the total time $T=<\tau> N$ on which one investigates
the recurrences of a trajectory, we have
$P(\tau) = <\tau> N_{\tau}/T \sim <\tau> \mu(\tau)/\tau$
where, due to ergodicity of motion 
the measure (probability to stay) is proportional to the ratio of 
time the trajectory spends in the region $(T_\tau \sim \tau N_\tau)$ 
to the total time $T \;\;\; (\mu(\tau) \sim T_\tau/T)$. 
Inside the sticking region
the dynamical variables are correlated so that $C(\tau) \sim \mu(\tau)$
\cite{kiev,chsh,dubna,chispan}. Since the correlations are
directly related to a diffusion rate $(D_c \sim \int C d\tau)$
the exponent $p<2$ can lead to a superdiffusive dynamics 
\cite{chsh,dubna,chispan,chilos}. 
For the standard map such a behavior was indeed
observed in \cite{chsh,dubna,chizetp}. All this show that the asymptotic decay
of Poincar\'e recurrences is a cornerstone statistical problem 
of two-dimensional maps.

To understand the asymptotic properties of $P(\tau)$ we used for the first time
a new approach based on a direct computation of 
{\em exit} times from a vicinity of 
the critical golden curve in the standard map
\begin{equation}
{\bar y} = y - K/(2 \pi) \sin (2\pi x) ~~,~~ 
{\bar x}= x + {\bar y}~~~\hbox{mod 1}
\label{stmap}
\end{equation}
with parameter $K=K_g$. The properties of this curve
had been studied in great detail \cite{mackay}. In particular,
the positions of unstable fixed points of resonances $p_n/q_n$
are known with high precision \cite{mackay}. To determine the exit time 
$\tau_n$ from 
the scale $q_n$ we placed 100 orbits in a very close vicinity of
an unstable fixed point and computed the average exit
time. For each orbit the exit time is determined as
a time after which the orbit crosses the exit line.
The exit line was fixed as  $y=1$ for the orbits from the side of the
main resonance $q=1$ or as $y=0.5+a \sin(2\pi x)$ for the orbits
from other side of the critical curve with $q=2$.
In the later case the exit line was drown in such a way to cross the
two unstable points of resonance $q=2\; (a=0.0773...)$.
This allowed to take into account the deformation of the $q=2$ resonances.
The average exit time $\tau_n$ from a given scale $q_n$
is related to the distance of this resonance from the
curve and is proportional to 
this distance (measure of chaos)
$\mu_n =  |r_g - r_n| \approx 1/\sqrt{5}q_n^2$
divided by the local diffusion rate 
$D_n: \;\; \tau_n \sim \mu_n/D_n \sim q_n$.
This gives $\mu \sim C \sim 1/\tau^2$ and $p=3$.
The numerical data for dependence of 
$\mu_n$ on $\tau_n$ is shown in Fig.1. From both sides of the $r_g$ curve
the exit times converge to the asymptotic dependence
\begin{equation}
\mu_n = (\tau_g/\tau_n)^2/\sqrt{5},~~ \tau_n =  \tau_g  ~ q_n,~~ 
\tau_g = 2.11 \cdot 10^5
\label{exit}
\end{equation}

This dependence corresponds to the 
scaling near the critical curve \cite{mackay,chispan}.
Indeed, the local diffusion rate in $y$ on the scale $q_n$
is $D_n \approx A D_0/q_n^5$, where $D_0=K^2/8\pi^2$ is the quasilinear
diffusion rate \cite{lib} and $A \approx 0.0066$ is a numerical constant
which is quite small due to a slow diffusion inside the separatrix layer 
\cite{ruffo}.
As a result the sum of transition times 
between the two scales from $r_n$ to $r_{n-2}$
is approximately equal to the total exit time:
$\tau_n \approx \sum_n |r_n - r_{n-2}|^2/D_n \approx 1.4 \cdot 10^5 q_n$.
This estimate gives the value of $\tau_g$ close to (\ref{exit}) and
allows to understand the physical origin of its large
numerical value. It is interesting to note that the data
of Fig.1 show that the convergence of $\tau_n$ to its 
asymptotic value can be satisfactory described as 
$|\tau_n/ q_n \tau_g - 1| \propto 1/q_n$. This indicates a
certain similarity between the ratio $\tau_n/q_n \tau_g$ and the residue
$R_n$ for periodic orbits which converges to its 
critical value in a similar way \cite{mackay}.
Physical reason of  this similarity is the following: $R_n$ is related to 
the orbit stability and the larger it is more rapidly 
the orbit escapes from the scale $q_n$. Due to that for 
odd $n (q_n=1,3,8,...)$ the time $\tau_n$ is smaller than
the asymptotic expression (\ref{exit})  ($R_n > R_{cr}=0.250...$
\cite{mackay}) while for even $n$ $(q_n=2,5,13,...)$
it is larger than (\ref{exit}) ($R_n < R_{cr}$ \cite{mackay}).
Due to universality of the critical golden curve structure
it is natural to expect that the relation (\ref{exit})
and the time $\tau_g$ are universal for all
area-preserving maps as well as $R_{cr}$ (note that
$q_n$ is the period of unstable periodic orbit on the scale $n$).

The relation (\ref{exit}) determines
the measure of chaotic region $\mu \sim \mu_n$ 
at which a trajectory is stuck
for a time $\tau \sim \tau_n$. Then, according to (\ref{corel}) the exponent 
of Poincar\'e recurrences is $p=3$ and correlations $C(\tau) \sim \mu$
decay as inverse square of time. However, this asymptotic decay starts in fact 
only after a very long time $\tau > 10^6$ due to the large value of $\tau_g$.
This strong delay of asymptotic behavior is responsible for the
nonuniversal decay observed for $P(\tau)$ in \cite{chsh,dubna}
at $K=K_g$. Indeed, for $\tau < \tau_g$ a trajectory doesn't feel
the border and $\mu$ remains approximately constant giving
$p=1$ that had been seen in \cite{chsh,dubna} (see Fig.2). However, to observe
the theoretical exponent $p=3$ one should go to longer times.
To check these theoretical expectations we made extensive numerical
simulations of $P(\tau)$ at $K=K_g$ with recurrences on the
exits lines defined above on the both sides of the critical $r_g$ curve. 
The results are presented in Fig.2 and show a clear change in 
decay of $P(\tau)$ for $\tau > 10^5$ (side $q=1$) and  $\tau > 10^7$
(side $q=2$). To check the relation (\ref{corel}) we 
computed $P(\tau)$ from the data of Fig.1 taking 
the recurrence time $\tau=2\tau_n$ and $P(\tau)=B_q <\tau>_q \mu_n/\tau$
where $q=1,2$ mark the side of critical curve. With the average
recurrence time $<\tau>_1 \approx 24.5$ (or $<\tau>_2 \approx 61$)
and an arbitrary constant $B_1=2.0$ (or $B_2=8.2$)  
the data from Fig.1 describe the variation of $P(\tau)$ 
in the interval of 6 (4) orders of magnitude. This gives additional
support for the theoretical exponent $p=3$.

Another check of the relation (\ref{corel}) was done by computing
the diffusion rate in phase $\bar{z}=z+(\bar{y}+y - 2 z_q)/2$
with $z_q=0 (q=1)$ and $z_q=1/2 (q=2)$. Similar approach was
used in \cite{chizetp}. The diffusion rate is $D_c= (\Delta z)^2 / \Delta t$
and its dependence on time is determined by the 
decay of correlation function of $y(t)$. According to (\ref{corel})
we have  $D_c (\tau) = D_{cq} G_q \int \tau P(\tau)/<\tau> d \tau = 
D_{cq} \tilde{G_q} \int C(\tau) d \tau $ where
$D_{cq} =|r_g - r_q|^2/3 = 0.049 (0.0046)$ is the quasilinear diffusion rate 
\cite{lib} for $q=1 \;(2)$ side
and $G_q,  \tilde{G_q}$ are some constants. The correlation $C(\tau)$ was
computed from the linear interpolation of data in Fig.1.
In addition we took that $C(2 \tau_n) = \mu_n/\mu_{n1}$
where $\mu_{n1}$ is the value of $\mu_n$ at the first scale on each side of 
the critical curve with $q=3,5$. 
In this way  $C(\tau)$ remains constant up to the first exit
time $\tau_{n1}$ $(C(0)=C(\tau_{n1})=1)$ that corresponds to the fact
that $P(\tau) \sim 1/\tau$ for $\tau<\tau_{n1} \sim \tau_g$.
For $p=3$ the rate $D_c$ should be finite.
The value of $D_c$ was computed for
100 orbits initially located near unstable fixed point of period
$q=1,2$. The diffusion rate dependence on $\tau$, and comparison
with its computation from $P(\tau)$ and exit times $\tau_n=\tau/2$
via the above integral relation, are shown in Fig.3. Both methods
give a good agreement with $D_c(\tau)$, especially in the case of $P(\tau)$,
confirming (\ref{corel}). The constants are $G_1 \approx G_2 \approx 2$,
$\tilde{G_1} \approx 0.3, \tilde{G_2} \approx 0.6$.
According to the above values of coefficients $B_q$ the ratio
$\tilde{G_q}/G_q$ should be approximately 2 times smaller.
This may be due to approximate scheme used to relate
$C(\tau)$ with $\mu(\tau)$. For $\tau > 10^7$ the asymptotic value
$p=3$ leads to a saturation of $D_c$ growth in time.
Even if the asymptotic diffusion rate is constant $D_c = D_c(\infty)$
the distribution function is non-gaussian since the higher
moments diverge.
For smaller $\tau$ the diffusion rate $D_c$ grows approximately linearly 
that corresponds to an intermediate value of $p \approx 1$. 
This intermediate slow decay is responsible for the enormously large
ratio of the asymptotic diffusion rate to its quasilinear value:
$D_c(\infty)/D_{cq} \approx 3 \cdot 10^5 (q=1); \;\; 10^7 (q=2)$.

The ensemble of data in Figs. 1- 3 shows that the asymptotic decay
of Poincar\'e recurrences and correlations is determined by the 
universal structure in the vicinity of the critical golden curve,
and the contribution of boundaries of other internal islands of stability
is not significant at variance with \cite{ott}. The hypothesis of
dynamical disconnection of scales \cite{dubna}
is also ruled out.
Our results are in agreement with previous
numerical observations indicated that long recurrences 
are related to orbits being very close to $r_g$ curve \cite{chsh,dubna}.
It is interesting to ask the question how  the value
of the exponent $p=3$ would be modified for a case of main border curve
$r_b = [a_1,a_2,...,a_i, ...]$
with non-golden continued fraction expansion. The numerical data
for border invariant curves obtained in \cite{stark} show that the elements
$a_i$ mainly take the values $1,2,3,4$ and the probability to find
$a_i > 4$ is rather small. For such bounded $a_i$ values
the general resonant approach developed in \cite{chispan,chilos,dubna}
still show that the diffusion rate near the border
scales as $D_n \propto 1/q_n^5$ and therefore the exit time will scale
as $\tau_n \sim q_n$ giving the exponent $p=3$. Due to similarity
between $\tau_n/q_n$ and the residue $R_n$ discussed above
we can expect that for a typical $r_b$ the ratio $\tau_n/q_n$
will not converge to a constant as it was for $r_b=r_g$ 
but will oscillate in a bounded interval similarly to 
oscillations of $R_n$ (see e.g. \cite{dubna}).
Due to above reasons
we can expect that even in the general case of non-golden border
invariant curves the average asymptotic universal exponent  is $p=3$.

If the exponent $p=3$ then it is natural to ask why 
the previous computations of different groups were giving
$p \approx 1.5$. Our explanation is based on the following arguments.
First, even for the critical golden curve the asymptotic
regime starts after very long time which is determined by first resonance
scales. The first scales are not universal that explains why
$p$ was varying from map to map. If the border curve
is non-golden then the ratio $\tau_n/q_n$ should oscillate with $n$ and
the asymptotic regime will appear even later than for $r_b = r_g$.
Also on the first scales a given map 
can be often locally approximated by the standard map with 
$K \approx K_{cr} + (r - r_b) d f(r)/ d r $,
where $f$ is some smooth function of rotation number \cite{chirikov}.
A typical example is the separatrix map \cite{chirikov,chsh,dubna}.
In this case at a given $r_n$ the local order parameter is supercritical
with $K - K_{cr} \propto |r_n - r_b| \propto 1/q_n^2$.
This scaling is different from the asymptotic one with
$K - K_{cr} \propto 1/q_n \propto |r_n - r_b|^{1/2}$ 
\cite{mackay,dubna,murray} and can give very long exit time
for first scales. Indeed, in the standard map
with $K>K_g$ the transition time from $y=0$ to $y=1$
is proportional to $1/|K - K_g|^3$ \cite{chirikov,meiss}
that will give exit time $\tau_n \sim  /|K - K_g|^3  \sim q_n^6$.
According to (\ref{corel}) this will give
$p=4/3$ that is not far from the average $p \approx 1.5$.
In addition, when being close to the critical curve,
as in the standard map with $K=K_g$,
one should still wait a long time $\tau_g$ to reach the asymptotic
exponent $p=3$.

In conclusion, we showed that in the case of critical golden curve
the asymptotic exponent for decay of Poincar\'e recurrences is $p=3$.
This implies that the correlation integral
converges and the diffusion rate produced by such dynamical chaos is
finite. However, the higher moments of distribution will diverge.
We argue that the asymptotic exponent should remain the same 
also in the case of a typical border invariant curve.

\begin{figure}
\vspace{4mm}                                 
\caption{
Dependence of exit time $\tau_n$ from the scale $r_n$
on a distance (chaos measure) from the critical golden curve
$\mu_n =|r_g - r_n|$ for $q_n = 3, 8, ... , 6765$
(black circles) and $q_n = 5, 13, ... , 4181$
(open circles). The straight line shows asymptotic 
behavior (3). Error bars are less than the symbol size.
Logarithms are decimal in Figs. 1 - 3.
}
\end{figure}

\begin{figure}
\vspace{4mm}
\caption{ 
Poincar\protect{\'e} recurrences in the standard map at $K=K_g$ from the
side of resonance $q=2$ (upper full curve) and
$q=1$ (lower full curve, shifted down by 3 for clarity).
Open and full circles show the values of $P(\tau)$ recalculated from
the data of Fig.1 (see text). The dashed straight lines mark the
asymptotic decay with theoretical exponent $p=3$;
slope $p=1$ is shown by dashed-dotted line.
Data for $P(\tau)$ are obtained from 10 orbits of length $10^{11}$.
}
\end{figure}

\begin{figure}
\vspace{4mm}                          
\caption{
Dependence of diffusion rate $D_c$ on time (full curves)
compared with its values computed from the Poincar\protect{\'e}
recurrences of Fig.2 (dashes curves) and exit times of 
Fig.1 (open and full circles) according to relation (1)
(see text). Lower curves and circles are for $q=1$ side,
while the upper ones are for $q=2$ (shifted up by 4 
for clarity). For clarity, all circles are shifted up by 0.3
from their optimal positions given by coefficients $\tilde{G_q}$
(see text).
}
\end{figure}

\end{multicols}

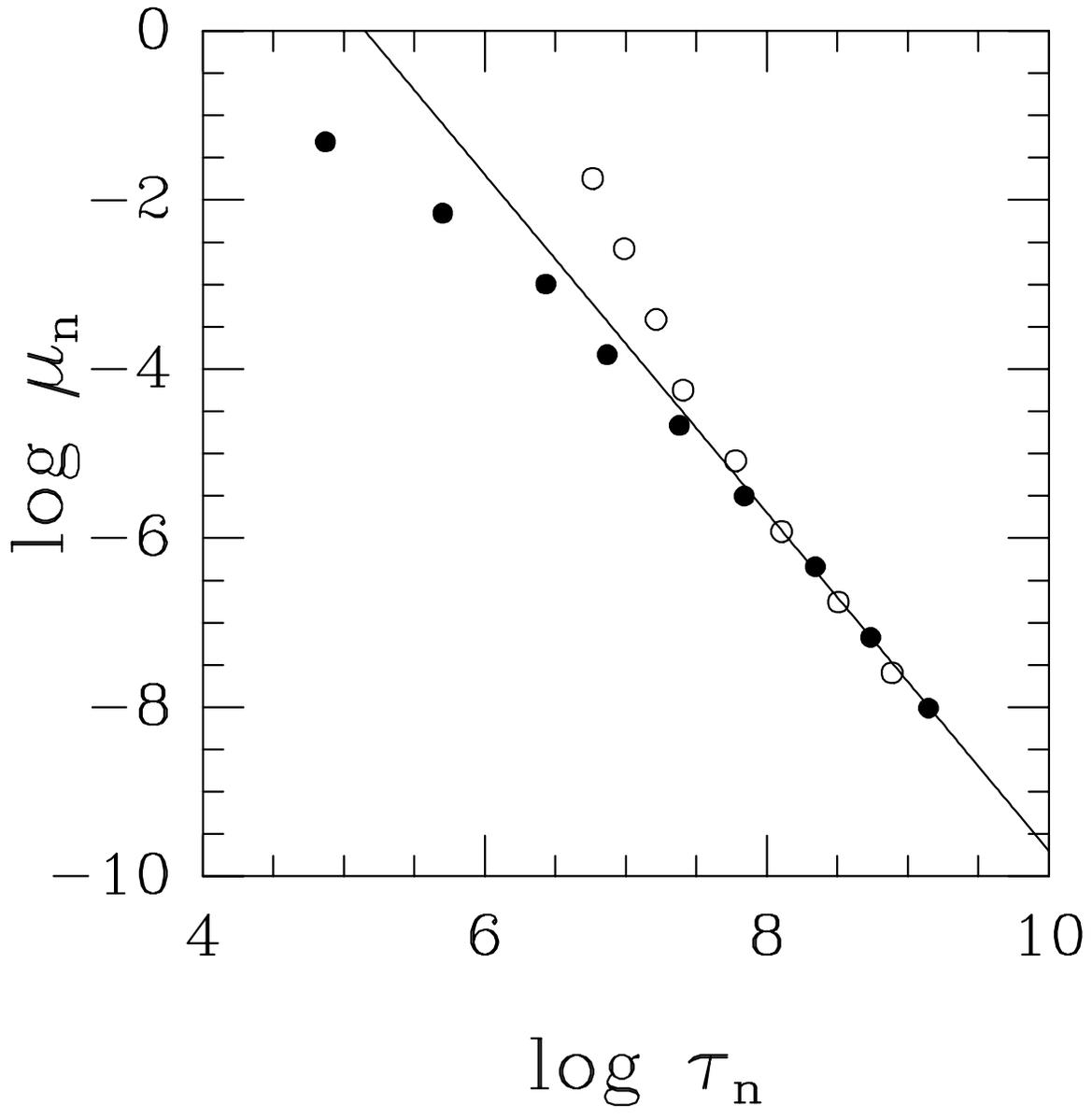
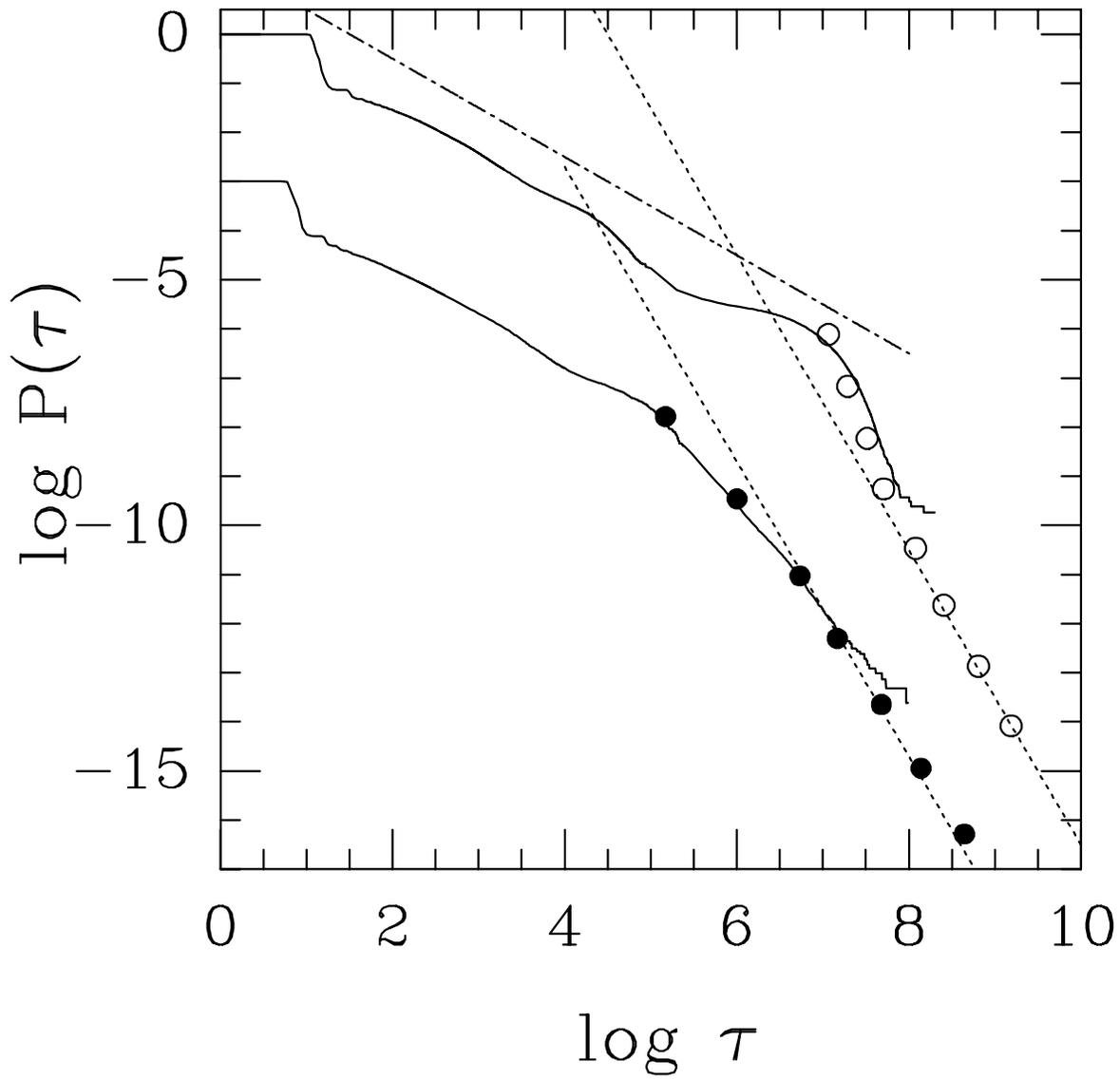
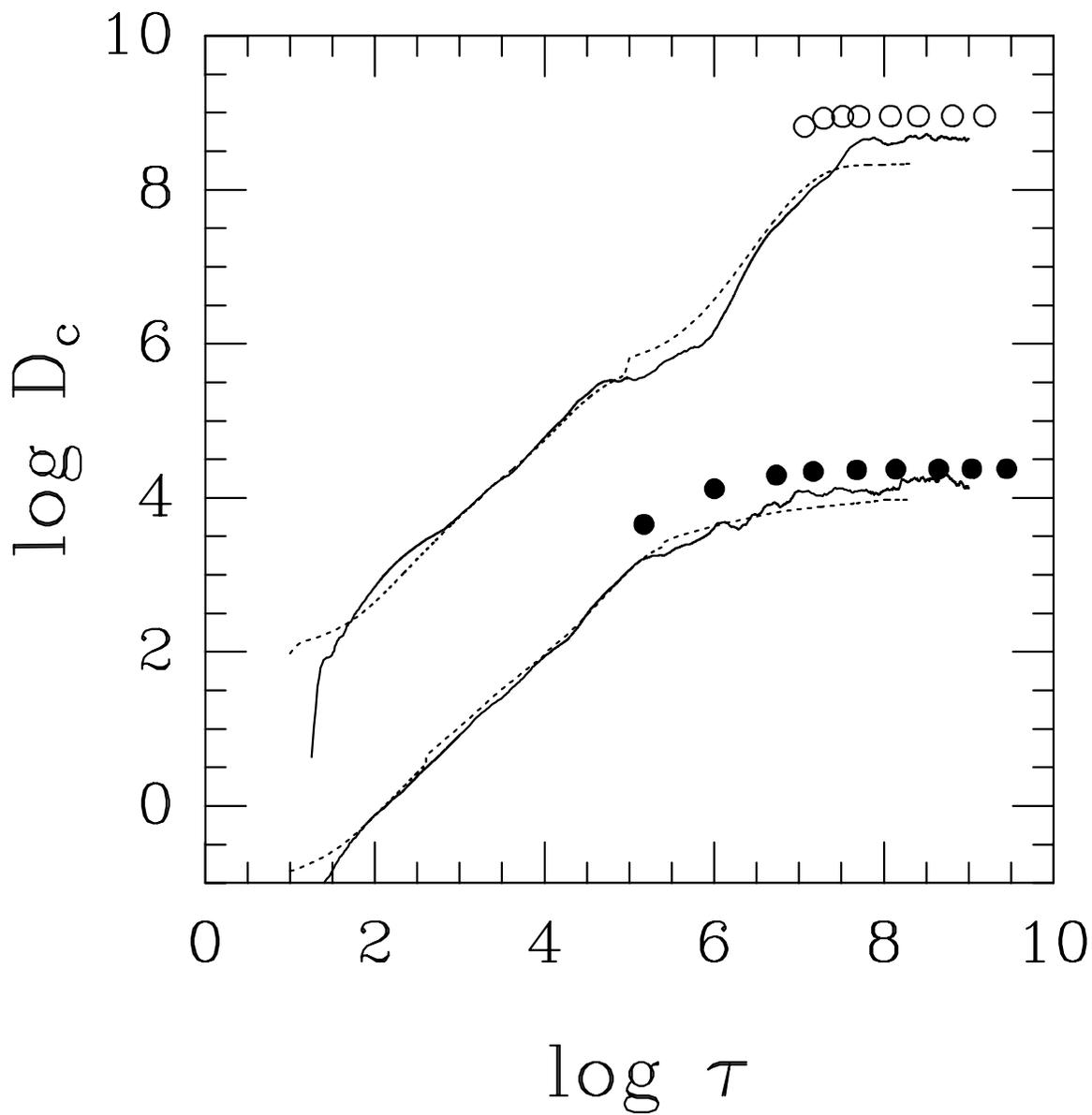
\end{document}